\documentclass[reprint,twocolumn,aps,prb,superscriptaddress]{revtex4-2}
\usepackage[T1]{fontenc}
\usepackage[utf8]{inputenc}
\usepackage{lmodern}
\usepackage{graphicx}
\usepackage{amsmath}
\usepackage{xcolor}
\usepackage{float}
\usepackage{siunitx}
\usepackage{hyperref}
\usepackage{braket}
\usepackage{amssymb,amsmath,amsthm, array}
\usepackage{mdframed}
\usepackage{systeme,mathtools}
\usepackage{yfonts}
\usepackage{comment}
\usepackage{bm}

\DeclareUnicodeCharacter{0301}{-}

\usepackage{tikz}
\usepackage{pgfplots}
\pgfplotsset{compat=1.18}
\definecolor{tealc}{HTML}{0F6E56}
\definecolor{coralc}{HTML}{B0451D}
\definecolor{grayc}{HTML}{5F5E5A}

\usepackage[caption=false]{subfig}

\begin{document}

\renewcommand{\vec}[1]{\mathbf{#1}}
\newcommand{\ii}{\mathrm{i}}
\def\ya#1{{\color{orange}{#1}}}

\title{Supersymmetric Neuromorphic FitzHugh--Nagumo model}

\author{D. Maroulakos}
\affiliation{Doctoral School at the  University of Rzesz\'ow, 35-317 Rzesz\'ow, Poland}

\author{A. Wal}
\affiliation{Institute of Physics, Faculty of Exact and Technical Sciences, University of Rzesz\'ow, 35-317 Rzesz\'ow, Poland}

\author{S. K. Mishra}
\affiliation{Department of Physics, Indian Institute of Technology (Banaras Hindu University) Varanasi - 221005, India}

\author{Z. Toklikishvili}
\affiliation{Faculty of Exact and Natural Sciences, Tbilisi State University, Chavchavadze av.3, 0128 Tbilisi, Georgia}

\author{L. Chotorlishvili}
\affiliation{Department of Physics and Medical Engineering, Rzesz\'ow University of Technology, 35-959 Rzesz\'ow, Poland}

\date{\today}

\begin{abstract}

We develop a supersymmetric spectral description of stochastic FitzHugh--Nagumo dynamics motivated by an effective antiferromagnetic neuromorphic model. A local normal-form reduction of the adopted tangential N\'eel-vector dynamics relates magnetic control parameters to fast activation and slow recovery coordinates. Freezing the recovery variable yields a one-dimensional Fokker--Planck equation with a quartic potential, whose similarity-transformed generator admits an exact supersymmetric
factorization. The first nonzero relaxation eigenvalue is determined by the ground state of the partner operator; in the weak-noise bistable regime, it is approximated by the sum of the two interwell Kramers rates. Projecting the similarity-transformed probability density onto the two lowest modes defines a normalized auxiliary two-state amplitude. We derive its localized-basis
coherence in terms of both the time-dependent basin populations and their equilibrium weights. Its maximum corresponds to equal normalized amplitude weights, which generally differ from equal basin populations. The fixed-recovery example exhibits a pronounced coherence pulse during monotonic relaxation without a spike. Conversely, the illustrated deterministic spiking trajectory repeatedly leaves the bistable domain and passes through regions with insufficient barrier separation, restricting the validity of the localized two-state approximation. These results connect stochastic relaxation spectra to an occupation-based coherence diagnostic and delimit its application to neuromorphic spiking. The diagnostic characterizes an auxiliary amplitude representation of classical stochastic dynamics and does not by itself establish microscopic quantum coherence.

\end{abstract}

\maketitle

\section{Introduction}

The FitzHugh--Nagumo (FHN) model is a well-known paradigmatic simplification of the four‑dimensional Hodgkin–Huxley equations~\cite{HodgkinHuxley1952}, and it describes the generation and propagation of action potentials in nerve axons. In the early 1960s, Richard FitzHugh~\cite{FitzHugh1961} and, independently, Jin‑ichi Nagumo and co‑workers~\cite{Nagumo1962} reduced the Hodgkin–Huxley dynamics to a two‑variable system that retains the essential features of neuronal excitability: a fast activation variable $v$ (mimicking the membrane potential) and a slow recovery variable $w$ (representing the gating of ion channels). The model is written as
\begin{align}
\frac{dv}{dt} &= v - \frac{v^{3}}{3} - w + I_{\text{ext}},\\[2pt]
\frac{dw}{dt} &= \varepsilon\,(v + a - b\,w.)\nonumber
\end{align}
Here, the small parameter $\varepsilon \ll 1$ leads to the separation of time scales, and $a$, $b$ are positive constants. The cubic nonlinearity $v - v^{3}/3$ is related to the fast inward sodium current, while the linear recovery term mimics the slower outward potassium current. The external stimulus $I_{\text{ext}}$ acts as a control parameter. The FHN system is a classic example of a slow–fast excitable system~\cite{FitzHugh1969, Rinzel1985}. Its phase‑plane geometry is conveniently analyzed through nullclines: the $v$‑nullcline (a cubic curve) and the $w$‑nullcline (a straight line). Depending on the position of the intersection, the system may be excitable, oscillatory, or bistable. For a rest state lying on the left branch of the cubic, a sufficiently large perturbation $I_{\text{ext}}$ pushes the trajectory beyond the middle branch, triggering a large‑amplitude excursion—the analog of an action potential—before returning to rest. This threshold behavior is the hallmark of excitability and is thoroughly explained in textbooks on mathematical physiology~\cite{KeenerSneyd1998} and mathematical biology~\cite{Murray2002}. Because of its simplicity and rich dynamics, the FHN model has become a cornerstone of computational neuroscience. It serves as the prototypical type~II neuron in Izhikevich’s classification of bursting and spiking neurons~\cite{Izhikevich2007}. The model also forms the basis for reaction–diffusion systems that describe propagating pulses and spiral waves in cardiac tissue and chemical media~\cite{Winfree2001}. Moreover, the FitzHugh–Nagumo reduction inspired many subsequent two‑variable models, such as the Morris–Lecar model~\cite{MorrisLecar1981}, which uses more realistic ionic current expressions while retaining the same mathematical skeleton. Today, the FHN equations remain a standard tool for teaching nonlinear dynamics, pattern formation, and the principles of biological oscillator theory. In the recent work  \cite{maroulakos2026hopf} we analyzed an antiferromagnetic spin-Hall nano-oscillator in the overdamped regime and derived a reduced set of equations for the Néel-vector dynamics constrained to the unit sphere. We proved that the projected dynamics of the Néel-vector can be transformed into a local FitzHugh-Nagumo normal form. Note that antiferromagnets are especially attractive in this context because they combine ultrafast dynamics, robustness against stray fields, and compatibility with spin-orbit torque and spin-pumping readout. For an antiferromagnet/heavy-metal bilayer, the charge current in the heavy metal generates a spin current through the spin Hall effect \cite{Baltz2018,Khymyn2018,Grollier2020}. Near a nonzero operating point, the local projected equations can be transformed into a canonical FHN normal form. This provides a direct connection between magnetic material parameters and the effective FHN parameters. A separate question is whether such an effective FHN model can be mapped into a Schr\"odinger equation and to qubit states. The answer requires two distinct but compatible steps. In the present work, we prove that the stochastic FHN dynamics can be rewritten in a Schr\"odinger-like form by using the Nelson map between Brownian stochastic mechanics and wave mechanics. In this description, the wave function lives in the continuous FHN state space. Due to the vital interest in neuromorphic synthetic AFM systems and the FHN model, we believe that mapping of the neuromorphic FHN model into the effective quantum qubit definitely deserves attention, and we dedicate our work to this problem.

\section{Model}

The spin current acts on the N\'eel vector $\bm I=\bm m_1-\bm m_2$, where $\mathbf m_{1,2}$ are the two sublattice magnetizations. In the overdamped regime, and for spin polarization along the easy axis, the projected N\'eel-vector dynamics is constrained to the unit sphere, $|\bm I|=1$. We consider a normalized N\'eel vector
\begin{equation}
    \bm I=(I_x,I_y,I_z), \qquad I_x^2+I_y^2+I_z^2=1.
\end{equation}
To proceed further, we introduce the action of the system:
\begin{eqnarray}\label{the action}
S=\int\mathcal L \left(\bm I, \dot{\bm I}\right)dt.
\end{eqnarray}
The action introduced here leads to the \textit{Euler-Lagrange-Rayleigh} equation projected on the sphere and tangential dynamics
\begin{eqnarray}\label{equations of Lagrange}
&& \bm I\times\frac{d}{dt}\frac{\partial\mathcal{L}}{\partial\dot{\textbf{I}}}=\bm I\times\frac{\partial\mathcal{L}}{\partial\textbf{I}}-\bm I\times\frac{\partial\mathcal{R}}{\partial\dot{\textbf{I}}},
\end{eqnarray}
where $\bm{I}=(I_x, I_y, I_z)$.  
Transition to the dimensionless equations can be done through the transformations: $\tau_m\equiv\omega_{\rm ex}\tau_m$, $\omega_0\equiv \omega_0/\omega_{\rm ex}$, $\omega_{\rm ip}\equiv \omega_{\rm ip}/\omega_{\rm ex}$, $\tau\equiv\tau/\omega_{\rm ex}$. Then the dimensionless Lagrange and Rayleigh functions read
\begin{eqnarray}\label{2appendix equations of Lagrange}
&& \mathcal L=\frac{1}{2}\dot{\bm I}^{\,2}-\frac{1}{2}
\left[\omega_0^2 (I_x^2 + I_y^2)+\omega_{\rm ip}^2 I_y^2
\right],\nonumber\\
&&\mathcal R=\frac{\alpha }{2}\dot{\bm I}^{\,2}+\tau\,\bm p \cdot (\bm I \times \dot{\bm I}),\nonumber\\
&&w_a=\frac{1}{2}\left[\omega_0^2 (I_x^2 + I_y^2)+\omega_{\rm ip}^2 I_y^2\right].
\end{eqnarray}
Here $\alpha$ is the Gilbert damping, $M_s$ is the sublattice magnetization, $A$ is the exchange stiffness, $H_{\mathrm{ex}}$ is the exchange field, $K$ is the easy-axis anisotropy constant, while  $K_{\mathrm{ip}}$ corresponds to the in-plane (secondary) anisotropy. The exchange frequency is described by $\omega_{ex}$. The easy-axis magnon gap: $\omega_0=\gamma\sqrt{\frac{H_{\rm ex}K}{M_s}}$, $\omega_{\rm ip}=\gamma\sqrt{\frac{H_{\rm ex}K_{\rm ip}}{M_s}}$ and the torque coefficient $\tau=\sigma j$.  

Our interest concerns the constrained Euler--Lagrange--Rayleigh (ELR) equation and a metric tangent force $\mathbf F_{\mathrm{ELR}}$ on the unit sphere. The operation $\mathcal J_{\bm I}\bm A=\mathbf A\times\bm I$ defines a rotated effective force. On tangent vectors $\mathcal J_{\bm I}^{\,2}=-P_{\bm I}$, so that $\mathcal J_{\mathbf I}$ acts as a rotation by $\pi/2$. We adopt $\mathbf F_T=\mathcal J_{\mathbf I}\mathbf F_{\mathrm{ELR}}$ as a minimal symmetry-consistent effective model for this rotational sector. Taking into account  
$\frac{\partial \mathcal L}{\partial \dot{\bm I}}=\dot{\bm I}$, $\frac{d}{dt}\frac{\partial \mathcal L}{\partial \dot{\bm I}}=\ddot{\bm I}$ and the scalar-triple-product identity
$\mathbf p\cdot\bigl(\bm I\times\dot{\bm I}\bigr)=\dot{\bm I}\cdot
\bigl(\mathbf p\times\bm I\bigr)$ we obtain explicit expression of the anisotropy field
\begin{eqnarray}\label{2the geometrical constraint}
&&\mathbf H_{\mathrm{an}}=\frac{\partial \mathcal L}{\partial \bm I}=
-\nabla_{\bm I} w_a=
\begin{pmatrix}
-\omega_0^2 I_x\\
-\bigl(\omega_0^2+\omega_{\mathrm{ip}}^2\bigr)I_y\\
0
\end{pmatrix},\\ 
&&\frac{\partial \mathcal R}{\partial \dot{\bm I}}=\alpha\dot{\bm I}
+\tau\bigl(\mathbf p\times\bm I\bigr).\nonumber
\end{eqnarray}
Substitution of Eq.(\ref{2the geometrical constraint}) into Eq.(\ref{equations of Lagrange})
leads to the following result 
\begin{eqnarray}\label{3the geometrical constraint}
&&\bm I\times\left[\ddot{\bm I}-\mathbf H_{\mathrm{an}}+\alpha\dot{\bm I}
+\tau\bigl(\mathbf p\times\bm I\bigr)\right]=\mathbf 0.
\end{eqnarray}
Applying $-\bm I\times$ extracts the metric tangent part. Using the orthogonal projector $P_{\bm I}$: $P_{\bm I}\mathbf A=\mathbf A-\bm I\bigl(\bm I\cdot\mathbf A\bigr)=
\bm I\times\bigl(\mathbf A\times\bm I\bigr)$, together with $P_{\bm I}\dot{\bm I}=\dot{\bm I}$ and $P_{\bm I}(\mathbf p\times\mathbf I)=\mathbf p\times\bm I$, we obtain
\begin{eqnarray}
P_{\bm I}\ddot{\bm I}
+
\alpha\dot{\bm I}
=
P_{\bm I}\mathbf H_{\mathrm{an}}
-
\tau\bigl(\mathbf p\times\bm I\bigr).
\end{eqnarray}
Applying $\mathcal J_{\bm I}$ to the tangent ELR force gives $\mathbf F_{\mathrm{T}}=\mathcal J_{\bm I}\mathbf F_{\mathrm{ELR}}$ with 
\begin{eqnarray}\label{1ELR}
&&\mathbf F_{\mathrm{ELR}}=P_{\bm I}\mathbf H_{\mathrm{an}}-\tau(\mathbf p\times\bm I)=
\mathbf H_{\mathrm{an}}-\\
&&(\bm I\cdot\mathbf H_{\mathrm{an}})\bm I-\tau(\mathbf p\times\bm I).\nonumber
\end{eqnarray}
In the pure tangent-torque model, we adopt
$\mathbf F_{\mathrm{T}}$ as the effective vector field:
\begin{eqnarray}\label{6ELR}
P_{\bm I}\ddot{\bm I}+\alpha\dot{\bm I}=
\mathbf F_{\mathrm{T}}.
\end{eqnarray}
The overdamped approximation neglects the
complete tangential inertial combination
$\mathsf P_{\bm I}\ddot{\bm I}$, while the normalization constraint remains exact. 
Neglecting the tangential inertial term but retaining the exact projection onto the unit
sphere, gives the first-order equation
\begin{equation}
 \alpha\dot{\bm I}=\bm F_{\mathrm T}(\bm I).
 \label{eq:overdamped-vector-new}
\end{equation}
In Cartesian components of the adopted overdamped tangent-torque model takes the form
\begin{eqnarray}\label{explicit in components}
&&\alpha\dot I_x=-\left(\omega_0^2+\omega_{ip}^2\right) I_y I_z+\tau\left[ p_x-(p_x I_x+p_z I_z)I_x\right] ,\nonumber\\
&&\alpha\dot I_y=\,\omega_0^2\, I_x I_z-\tau\,(p_x I_x+p_z I_z)\,I_y,\\
&&\alpha\dot I_z=\omega_{ip}^2\, I_x I_y+\tau\left[ p_z-(p_x I_x+p_z I_z)I_z\right].\nonumber
\end{eqnarray}
For spin polarization along the easy axis, $\mathbf p=p_z\hat z$, the overdamped projected equations can be written as \cite{maroulakos2026hopf}:
\begin{align}\label{overdamped}
\alpha \dot I_x &= -A I_y I_z-\tau p_z I_z I_x, \nonumber\\
\alpha \dot I_y &= B I_x I_z-\tau p_z I_z I_y, \\
\alpha \dot I_z &= C I_x I_y+\tau p_z(1-I_z^2),\nonumber 
\end{align}
where $\alpha$ is the Gilbert damping, $\tau$ is the spin-torque coefficient, and $A=\omega_0^2+\omega_{\rm ip}^2$, $B=\omega_0^2$, $C=\omega_{\rm ip}^2$ are determined by the easy-axis and in-plane anisotropy frequencies. The equations preserve the unit-sphere constraint and therefore describe tangential dynamics of the order parameter.

To obtain the local FHN structure, introduce a north-pole ansatz
\begin{equation}\label{variable v}
    v=I_x, \qquad w=1-I_z,
\end{equation}
so that
\begin{equation}\label{variable w}
    I_y=s\sqrt{2w-w^2-v^2}, \qquad s=\pm 1,
\end{equation}
with the validity condition $2w-w^2-v^2>0$. Derivation of the effective FHN model is cumbersome, for details we refer to the work \cite{maroulakos2026hopf}. Using Eq.(\ref{variable v}), Eq.(\ref{variable w}) the exact two-dimensional local dynamics has the form
\begin{align}
\dot v &= F(v,w)
=\frac{1-w}{\alpha}\left[-As\sqrt{2w-w^2-v^2}-\tau p_z v\right], \\
\dot w &= G(v,w)
=-\frac{1}{\alpha}\left[ Cvs\sqrt{2w-w^2-v^2}+\tau p_z(2w-w^2)\right].
\end{align}
Let $(v_\ast,w_\ast)$ be a local operating point and define
\begin{equation}\label{local_points}
    v=v_\ast+\xi, \qquad w=w_\ast+\eta.
\end{equation}
Expanding the fast equation to cubic order in $\xi$ and to first order in $\eta$ gives
\begin{equation}
    \dot\xi=f_0+f_1\xi+f_2\xi^2+f_3\xi^3+f_w\eta+\cdots.
\end{equation}
After the shift $\xi=x+h$, with $h=-f_2/(3f_3)$, the quadratic term is removed:
\begin{equation}
    \dot x=\tilde f_0+\tilde f_1x+f_3x^3+f_w\eta+\cdots,
\end{equation}
where
\begin{equation}
\tilde f_1=f_1-\frac{f_2^2}{3f_3},\qquad
\tilde f_0=f_0-\frac{f_1f_2}{3f_3}+\frac{2f_2^3}{27f_3^2}.
\end{equation}
If $\tilde f_1 f_3<0$, one may introduce the rescaled time $\theta=\tilde f_1 t$ and the normalized fast variable $V=\lambda x$, with
\begin{equation}
    \lambda^2=-\frac{3f_3}{\tilde f_1}.
\end{equation}
The recovery variable is chosen as
\begin{equation}
    W=-\frac{\lambda f_w}{\tilde f_1}\eta,
\end{equation}
and the local antiferromagnetic dynamics becomes
\begin{align}
\frac{dV}{d\theta} &= V-\frac{V^3}{3}-W+I_0, \\
\frac{dW}{d\theta} &= \epsilon(V+a-bW),\nonumber
\end{align}
where $I_0=\lambda\tilde f_0/\tilde f_1$ is an intrinsic effective bias generated by the magnetic normal-form transformation. An externally supplied spin-pumping input can be added as
\begin{equation}
    I_{\rm eff}(\theta)=I_0+\chi_{\rm sp}I_{\rm sp}^{\rm ext}(\theta),
\end{equation}
leading to
\begin{eqnarray}
&&\frac{dV}{d\theta} = V-\frac{V^3}{3}-W+I_{\rm eff}(\theta),\label{eq:fhn1}\\
&&\frac{dW}{d\theta} = \epsilon(V+a-bW). \nonumber
\end{eqnarray}
The inverse local map from the canonical FHN variables to the N\'eel-vector is
\begin{eqnarray}\label{eq:inversemap}
&&v(V)=v_\ast+h+\frac{V}{\lambda},\\
&& w(W)=w_\ast-\frac{\tilde f_1}{\lambda f_w}W.\nonumber
\end{eqnarray}
Equations (\ref{eq:fhn1})--(\ref{eq:inversemap}) are the central classical bridge: $V$ and $W$ are continuous coordinates encoding the local N\'eel-vector motion, not discrete qubit states.
\begin{table*}[ht]
\centering
\renewcommand{\arraystretch}{1.3}
\small
\begin{tabular}{c| c| c|c| c |c |c |c}
\hline
Case &
$\tau$ &
$p_z$ &
$v_*$ &
$w_*$ &
$a$ &
$b$ &
$\epsilon$ \\
\hline

1&
4.8000e-02 &
3.0900e-02 &
-1.6528e-01 &
3.0945e-01 &
1.2284e+00 &
1.3304e+00 &
5.2852e-02 \\

2&
2.0000e-03 &
9.0000e-01 &
-1.6699e-01 &
1.6864e+00 &
1.2869e+00 &
1.3536e+00 &
1.0917e-01 \\
\hline
\end{tabular}
\caption{Optimization results for parameter set $v_*$, $w_*$, given by Eq.(\ref{local_points}) and the FitzHugh-Nagumo neuromorphic map parameters $a$, $b$ and $\epsilon$. All parameters are obtained for $\tau$ and $p_z$ in Eq.(\ref{overdamped}).}
\label{tab:all_cases}
\end{table*}
Optimization results for parameter set $v_*$, $w_*$, given by Eq.(\ref{local_points}) and the FitzHugh-Nagumo neuromorphic map parameters $a$, $b$ and $\epsilon$ are presented in the Table \ref{tab:all_cases}.

\section{Schr\"odinger-like equation}

Thermal, electronic, and interfacial fluctuations can be included by promoting the deterministic FHN normal form to an It\^o stochastic differential equation:
\begin{align}
 dV_\theta &= \left[V_\theta-\frac{V_\theta^3}{3}-W_\theta+I_{\rm eff}(\theta)\right]d\theta
 +\sigma_V dB_V(\theta), \\
 dW_\theta &= \epsilon(V_\theta+a-bW_\theta)d\theta
 +\sigma_W dB_W(\theta),\nonumber
 \label{eq:sdefhn}
\end{align}
where $B_V$ and $B_W$ are independent Brownian motions. The corresponding probability density $P(V,W,\theta)$ satisfies Fokker-Planck equation \cite{Risken1989,CooperKhareSukhatme1995,Kramers1940,HanggiTalknerBorkovec1990,BaumgratzCramerPlenio2014}: 
\begin{eqnarray}\label{FP 1}
&&\frac{\partial P}{\partial\theta}=-\frac{\partial}{\partial V}\left(A_VP\right)-\frac{\partial}{\partial W}\left(A_WP\right)
+\frac{\sigma_V^2}{2}\frac{\partial^2 P}{\partial V^2}\nonumber\\
&&+\frac{\sigma_W^2}{2}\frac{\partial^2 P}{\partial W^2},
\end{eqnarray}
with drift components
\begin{eqnarray}
&&A_V=V-\frac{V^3}{3}-W+I_{\rm eff}(\theta),\\
&&A_W=\epsilon(V+a-bW).\nonumber
\end{eqnarray}
This density is the object that enters the Schr\"odinger-like and mixed-qubit descriptions below. The stochastic FHN system can be mapped to a wave-mechanical representation by writing the probability density $\psi_F=\psi_F(V,W,\theta)$ in polar form,
\begin{eqnarray}\label{1Nelson}
\psi_F(V,W,\theta)=\sqrt{P(V,W,\theta)}\exp\left[\frac{iS(V,W,\theta)}{\hat\hbar_{\rm AF}}\right].
    \label{eq:nelsonwave}
\end{eqnarray}
Here $S$ is an effective action field and $\hat\hbar_{\rm AF}$ is an effective stochastic action scale. In the simplest isotropic Nelson construction, $\hat\hbar_{\rm AF}=m_{\rm eff}\sigma$, where $m_{\rm eff}$ is an effective inertial or circuit parameter and $\sigma$ is the diffusion scale. For anisotropic noise, one may either rescale $V$ and $W$ to isotropic coordinates or introduce an effective mass/diffusion tensor.

To proceed further we utilize two different time scales of the problem. Namely we adopt the adiabatic approximation and tackle the slow variable $W$ as a adiabatic parameter. Then the modified Eq.(\ref{FP 1}) reads:
\begin{eqnarray}
&&\frac{\partial P_w(V,t)}{\partial t}=\frac{\partial}{\partial V}\left[\frac{\partial}{\partial V}+\beta U'_W(V)\right]P(V,t).
\end{eqnarray}
Here $t=\theta\sigma_v^2/2$, $\beta=2/\sigma_v^2$, and $U'_W(V)=-A_V$, so that
\begin{eqnarray}\label{eq:susy_fast_potential}
U_W(V)=\frac{V^4}{12}-\frac{V^2}{2}+(W-I)V . 
\end{eqnarray}

The nonlinear potential Eq.(\ref{eq:susy_fast_potential}) leads to the quantum parametric resonance in the vicinity of a separatrix solutions \cite{chotorlishvili2010quantum,PhysRevE.68.026216,PhysRevE.70.026219,PhysRevE.71.056211}. However, in this project we are interested in the lowest eigen states.

\section{Nelson map} 

For completeness, we spell out the Nelson stochastic map \cite{Nelson1966,Madelung1927} used in Eq.~(\ref{1Nelson}). Let
$X_\theta=(V_\theta,W_\theta)^T$, $\bm A=(A_V,A_W)^T$, and $\bm D=\mathrm{diag}(D_V,D_W)$, with $D_V=\sigma_V^2/2$ and
$D_W=\sigma_W^2/2$. The forward drift is $\bm b=\bm A$. For a smooth positive
density $P$, the backward drift is
\begin{eqnarray}\label{2Nelson}
\bm b_* = \bm b-2\bm D\nabla\ln P .
\end{eqnarray}
The Nelson osmotic and current velocities are therefore
\begin{eqnarray}\label{3Nelson}
&& \bm u={\bm b-\bm b_*\over2}=\bm D\nabla\ln P,\\
&&\bm v={\bm b+\bm b_*\over2}=\bm b-\bm D\nabla\ln P.\nonumber
\end{eqnarray}
With these definitions the Fokker--Planck equation becomes the continuity equation
\begin{eqnarray}\label{4Nelson}
\partial_\theta P+\nabla\cdot(P\bm v)=0.
\end{eqnarray}
If the current velocity is locally irrotational in the effective mass metric, one introduces an action field $S$ by
\begin{eqnarray}\label{5Nelson}
&&\bm v=\bm M^{-1}\nabla S, \qquad \bm M=\mathrm{diag}(m_V,m_W),\nonumber\\
&& D_i={\widehat\hbar_{\rm AF}\over 2m_i}.
\end{eqnarray}
For anisotropic noise this condition is equivalently obtained after rescaling the two coordinates, or by keeping the mass tensor $\bm M$. The Nelson wave amplitude is then Eq.(\ref{1Nelson}).
The mean acceleration is defined through the forward and backward derivatives
\begin{eqnarray}\label{7Nelson}
&&D_\pm f=\left(\partial_\theta+\bm b_\pm\cdot\nabla\pm \bm D:\nabla\nabla\right)f,\nonumber\\
&&\bm a_N={1\over2}(D_+D_-+D_-D_+)X .
\end{eqnarray}
Imposing $\bm M\bm a_N=-\nabla\Phi_N$ gives the quantum Hamilton--Jacobi equation
\begin{eqnarray}\label{8Nelson}
\partial_\theta S+{1\over2}\nabla S^T\bm M^{-1}\nabla S+
\Phi_N+Q_N=0,
\end{eqnarray}
where
\begin{eqnarray}\label{9Nelson}
Q_N=-\sum_{i=V,W}{\widehat\hbar_{\rm AF}^2\over2m_i}{\partial_i^2\sqrt P\over\sqrt P},
\end{eqnarray}
is the Nelson quantum potential. Equations (\ref{4Nelson}) and (\ref{8Nelson}) are equivalent to the Schr\"odinger-like equation $i=V,W$
\begin{eqnarray}\label{10Nelson}
i\widehat\hbar_{\rm AF}\,\partial_\theta\psi_F=\left[-\sum_{i}{\widehat\hbar_{\rm AF}^2\over2m_i}\partial_i^2+\Phi_N(V,W,\theta)\right]\psi_F .
\end{eqnarray}
Here $\Phi_N$ is the scalar potential that reproduces the prescribed stochastic drift through the Nelson equations. Since the full AFM--FHN drift is coupled and non-gradient, a single global scalar-potential representation need not exist. The map should therefore be understood locally, or after the slow--fast reduction used below. In the controlled fast-coordinate construction, the slow coordinate $W$ is frozen. With the sign convention used in the supersymmetric reduction,
\begin{eqnarray}\label{11Nelson}
&& A_V(V;W)=-U'_W(V),\\
&&U_W(V)={V^4\over12}-{V^2\over2}+(W-I)V.\nonumber
\end{eqnarray}
The one-dimensional Nelson velocities become
\begin{eqnarray}\label{12Nelson}
&&u_W=D_V\partial_V\ln P_W,\\
&&v_W=A_V-D_V\partial_V\ln P_W={1\over m_V}\partial_V S_W.\nonumber
\end{eqnarray}
Thus
\begin{eqnarray}\label{13Nelson}
&&\psi_W(V,\theta)=\sqrt{P_W(V,\theta)}\exp\left[{iS_W(V,\theta)\over\widehat\hbar_{\rm AF}}\right],
\end{eqnarray}
and
\begin{eqnarray}\label{14Nelson}
&&i\widehat\hbar_{\rm AF}\partial_\theta\psi_W=\left[-{\widehat\hbar_{\rm AF}^2\over2m_V}\partial_V^2+\Phi_W(V,\theta)\right]\psi_W.
\end{eqnarray}
The zero-current equilibrium condition $v_W=0$ gives
\begin{eqnarray}\label{15Nelson}
&&P_{\rm eq}(V|W)\propto\exp\left[-{U_W(V)\over D_V}\right]=\exp[-\beta U_W(V)],\nonumber\\
&&\beta={1\over D_V}={2\over\sigma_V^2}.
\end{eqnarray}
Independently of the Nelson representation, the frozen one-dimensional Fokker--Planck generator admits the similarity transformation and supersymmetric factorization constructed below.

\section{Supersymmetric theory}

The following full stochastic AFM--FHN model is two dimensional and non-gradient. Therefore, as discussed above, it cannot be represented exactly by a two-dimensional separable scalar-potential Hamiltonian. A controlled supersymmetric construction is obtained by using the slow--fast structure of the FHN equations. We freeze the slow recovery coordinate $W$ on the fast activation time scale and construct a one-dimensional Fokker--Planck operator for the fast coordinate $V$. Therefore, in the spirit of the Supersymmetric (SUSY) theory \cite{witten1981dynamical,parisi1982supersymmetric,PhysRevLett.52.1933,PhysRevB.102.024413}, the imaginary time Schr\"odinger equation reads 
\begin{eqnarray}\label{1SUSY}
\frac{\partial\chi_W}{\partial t}=-\hat H\chi_{W},
\end{eqnarray}
where the Hermitian, positive semidefinite Hamiltonian operator reads
\begin{eqnarray}\label{2SUSY}
\hat H=\mathcal{A}^\dag\mathcal{A},
\end{eqnarray}
with 
\begin{eqnarray}\label{3SUSY}
&&\mathcal{A}=\frac{1}{2}\beta U'(V)+\partial/\partial V,\\
&& \mathcal{A}^\dag=\frac{1}{2}\beta U'(V)-\partial/\partial V\nonumber.
\end{eqnarray}
The additive constant in $U_W$ is irrelevant. With this convention,
\begin{eqnarray}
&&\frac{\partial P_W}{\partial t}=\partial_V\left[\partial_V+\beta U_W'(V)\right]P_W .
    \label{eq:fp_standard_susy}
\end{eqnarray}
The zero-current stationary density is
\begin{eqnarray}\label{eq:peq_susy}
&&P_{\rm eq}(V|W)=\frac{1}{Z_W}\exp[-\beta U_W(V)],\\
&&Z_W=\int_{-\infty}^{\infty}e^{-\beta U_W(V)}\,dV.\nonumber 
\end{eqnarray}
Let us see how the general formalism works in our particular case. Before we note that the sign convention is important: if one instead writes $U'(V)=A_V$, then the signs in the supersymmetric operators must be reversed. The standard Bernstein--Brown-Witten factorization corresponds to the convention $U_W'=-A_V$. Introducing the similarity transformation
\begin{equation}\label{eq:susy_similarity}
    P_W(V,t)=e^{-\beta U_W(V)/2}\,\chi_W(V,t),
\end{equation}
Eq.~\eqref{eq:fp_standard_susy} becomes an imaginary-time Schrödinger equation
\begin{equation}
    \frac{\partial \chi_W}{\partial t} =-\hat H_B(W)\chi_W,
    \label{eq:imag_time_schrodinger_susy}
\end{equation}
where
\begin{equation}
    \hat H_B(W)=\mathcal A^\dagger\mathcal A .
\end{equation}
The first-order supersymmetric operators are
\begin{equation}
\mathcal A=\partial_V+\mathcal W_S(V;W),\qquad \mathcal A^\dagger =-\partial_V+\mathcal W_S(V;W),
    \label{eq:susy_A_ops}
\end{equation}
with superpotential
\begin{equation}
\mathcal W_S(V;W)=\frac{\beta}{2}U_W'(V)=\frac{\beta}{2}\left(\frac{V^3}{3}-V+W-I\right).
    \label{eq:superpotential}
\end{equation}
Consequently,
\begin{equation}
\hat H_B(W)=-\partial_V^2+\mathcal W_S^2(V;W)-\partial_V\mathcal W_S(V;W).
    \label{eq:HB_general}
\end{equation}
Since
\begin{equation}
    U_W''(V)=V^2-1,
\end{equation}
the bosonic Hamiltonian is explicitly
\begin{equation}
\hat H_B(W)=-\partial_V^2+\frac{\beta^2}{4}\left(\frac{V^3}{3}-V+W-I\right)^2-\frac{\beta}{2}\left(V^2-1\right).
\label{eq:HB_explicit}
\end{equation}
The supersymmetric partner Hamiltonian is
\begin{equation}
    \hat H_F(W)=\mathcal A\mathcal A^\dagger,
\end{equation}
namely
\begin{eqnarray}
&&\hat H_F(W)=-\partial_V^2+\frac{\beta^2}{4}\left(\frac{V^3}{3}-V+W-I\right)^2+\\
&&\frac{\beta}{2}\left(V^2-1\right).\nonumber
\label{eq:HF_explicit}
\end{eqnarray}
The full supersymmetric Hamiltonian may be written as
\begin{equation}
\hat H_{\rm SUSY}=\{Q,Q^\dagger\}=
 \begin{pmatrix}
        \hat H_B & 0\\
        0 & \hat H_F
    \end{pmatrix},
\end{equation}
with nilpotent supercharges
\begin{equation}
Q=\begin{pmatrix}
        0 & 0\\
        \mathcal A & 0
    \end{pmatrix},
    \qquad
    Q^\dagger=
    \begin{pmatrix}
        0 & \mathcal A^\dagger\\
        0 & 0
    \end{pmatrix},
    \qquad
    Q^2=(Q^\dagger)^2=0 .
\end{equation}
The bosonic ground state is annihilated by $\mathcal A$:
\begin{equation}
    \mathcal A\chi_{0,B}=0 .
\end{equation}
Therefore
\begin{equation}\label{eq:bosonic_ground_state}
\chi_{0,B}(V|W)=\frac{1}{\sqrt{Z_W}}\exp\left[-\frac{\beta}{2}U_W(V)\right],
\end{equation}
where $\lambda_{0,B}=0$. The original stationary probability density is recovered as
\begin{equation}
    P_{\rm eq}(V|W)=\chi_{0,B}^2(V|W).
\end{equation}
All nonzero eigenvalues occur in supersymmetric pairs:
\begin{equation}
    \lambda_{n,F}=\lambda_{n+1,B},
    \qquad n=0,1,2,\ldots .
\end{equation}
Thus the first nonzero relaxation eigenvalue of the Fokker--Planck problem is the ground-state energy of the fermionic partner Hamiltonian,
\begin{equation}
    \lambda_1\equiv \lambda_{1,B}=\lambda_{0,F} .
    \label{eq:lambda1_susy_pairing}
\end{equation}
The corresponding bosonic first excited state can be reconstructed from the fermionic ground state as
\begin{equation}\label{eigenfunction2}
    \chi_{1,B}(V|W)
    =\frac{1}{\sqrt{\lambda_1}}
    \mathcal A^\dagger\chi_{0,F}(V|W).
\end{equation}
This is the practical advantage of the supersymmetric construction: a difficult double-well bosonic relaxation problem is replaced by the ground-state problem of the partner Hamiltonian.
The potential in Eq.~\eqref{eq:susy_fast_potential} is bistable when
\begin{equation}
    \Delta\equiv W-I,
    \qquad
    |\Delta|<\frac{2}{3} .
\end{equation}
In this regime the extrema are the three real roots of
\begin{equation}
    U_W'(V)=\frac{V^3}{3}-V+\Delta=0 .
\end{equation}
Let these roots be ordered as
\begin{equation}
    V_L<V_S<V_R,
\end{equation}
where $V_L$ and $V_R$ are the two minima and $V_S$ is the barrier top. We define
\begin{eqnarray}
&&\kappa_L=U_W''(V_L)=V_L^2-1,\\
&&\kappa_R=U_W''(V_R)=V_R^2-1,\nonumber
\end{eqnarray}
with $\kappa_L,\kappa_R>0$, and
\begin{equation}
    \kappa_S=|U_W''(V_S)|=1-V_S^2>0 .
\end{equation}
The barrier heights are
\begin{eqnarray}
&&\Delta U_L=U_W(V_S)-U_W(V_L),\\
&&\Delta U_R=U_W(V_S)-U_W(V_R).\nonumber
\end{eqnarray}
In the weak-noise limit $\beta\gg 1$, the transition rates in the original FHN time $\theta$ are
\begin{equation}\label{eq:kramers_lr}
k_{L\rightarrow R}^{(\theta)}(W)\simeq\frac{\sqrt{\kappa_L\kappa_S}}{2\pi}\exp[-\beta\Delta U_L],
\end{equation}
\begin{equation}\label{eq:kramers_rl}
k_{R\rightarrow L}^{(\theta)}(W)\simeq\frac{\sqrt{\kappa_R\kappa_S}}{2\pi}\exp[-\beta\Delta U_R]. 
\end{equation}
Therefore the first nonzero Fokker--Planck eigenvalue is
\begin{equation}\label{eq:lambda1_theta}
    \lambda_1^{(\theta)}(W)
    \simeq
    k_{L\rightarrow R}^{(\theta)}(W)
    +k_{R\rightarrow L}^{(\theta)}(W) .
\end{equation}
Because the supersymmetric Hamiltonian was written in the rescaled time $t=D_V\theta$, the corresponding eigenvalue in the SUSY equation is
\begin{eqnarray}\label{eq:lambda1_t}
&&\lambda_1^{(t)}(W)=\beta\lambda_1^{(\theta)}(W)\simeq\frac{\beta}{2\pi}\bigg[\sqrt{\kappa_L\kappa_S}\,e^{-\beta\Delta U_L}+\nonumber\\
&&\sqrt{\kappa_R\kappa_S}\,e^{-\beta\Delta U_R}\bigg].
\end{eqnarray}
For the symmetric case $W=I$, one has
\begin{equation}
    V_L=-\sqrt{3},
    \qquad
    V_S=0,
    \qquad
    V_R=\sqrt{3},
\end{equation}
\begin{equation}
    \kappa_L=\kappa_R=2,
    \qquad
    \kappa_S=1,
    \qquad
    \Delta U_L=\Delta U_R=\frac{3}{4} .
\end{equation}
Thus
\begin{eqnarray}\label{eq:lambda1_symmetric}
&&\lambda_1^{(\theta)}\simeq\frac{\sqrt{2}}{\pi}\exp\left(-\frac{3\beta}{4}\right),\\
&&\lambda_1^{(t)}\simeq\frac{\beta\sqrt{2}}{\pi}\exp\left(-\frac{3\beta}{4}\right).\nonumber
\end{eqnarray}
This exponentially small eigenvalue controls the slow interwell relaxation of the noisy fast FHN coordinate and, in the excitable regime, the noise-assisted activation rate across the fast threshold.

The slow recovery variable can now be reintroduced by averaging over the conditional fast distribution. The adiabatic closure is
\begin{equation}
    \frac{dW}{d\theta}
    =\epsilon\left(\langle V\rangle_W+a-bW\right),
    \label{eq:slow_closure}
\end{equation}
where
\begin{equation}
    \langle V\rangle_W
    =\int_{-\infty}^{\infty}V P_W(V,\theta)\,dV .
\end{equation}
In local equilibrium,
\begin{equation}
    \langle V\rangle_W
    \simeq
    \frac{\int_{-\infty}^{\infty}V e^{-\beta U_W(V)}dV}
         {\int_{-\infty}^{\infty}e^{-\beta U_W(V)}dV} .
\end{equation}
In the weak-noise single-well limit, this reduces to
\begin{eqnarray}
&&\langle V\rangle_W\simeq V_m(W),\qquad V_m-\frac{V_m^3}{3}-W+I=0,\\
&&V_m^2>1.\nonumber
\end{eqnarray}
In the bistable regime, the conditional distribution can be approximated by a two-state mixture
\begin{equation}
    P_W(V,\theta)
    \simeq
    p_L(\theta)P_L(V|W)+p_R(\theta)P_R(V|W),
\end{equation}
where $P_L$ and $P_R$ are local Gaussian distributions around $V_L$ and $V_R$. The occupation probabilities obey
\begin{equation}
    \dot p_L=-k_{L\rightarrow R}^{(\theta)}p_L
    +k_{R\rightarrow L}^{(\theta)}p_R,
    \qquad
    p_R=1-p_L .
\end{equation}
The mean fast coordinate is then
\begin{equation}
    \langle V\rangle_W\simeq p_LV_L+p_RV_R .
\end{equation}
This provides a closed supersymmetric/adiabatic approximation to the stochastic AFM--FHN dynamics.

Finally, the same eigenfunctions define a natural two-level Hilbert-space truncation of the fast stochastic dynamics. We use
\begin{eqnarray}
&&\varphi_0(V;W)=\chi_{0,B}(V|W),\\
&&\varphi_1(V;W)=\chi_{1,B}(V|W),\nonumber
\end{eqnarray}
or, in the bistable regime, localized left and right wave packets $i=L,R$
\begin{eqnarray}\label{1solution}
\varphi_{i}(V;W)\simeq\left(\frac{\beta\kappa_{i}}{2\pi}\right)^{1/4}\exp\left[-\frac{\beta\kappa_{i}}{4}(V-V_{i})^2\right].
\end{eqnarray}
The consistency with the FHN kinetics holds under the condition
$\chi_W(V,\theta)\equiv\chi_W(V,t=D_V\theta)$, where
$D_V=\sigma_V^2/2=1/\beta$. Considering two-level Hilbert-space truncation we expand the states in the spectral basis $\{|0_F\rangle,|1_F\rangle\}$: 
\begin{eqnarray}\label{normalization1} 
&&|\chi_2(\theta;W)\rangle=\sum_{j=0}^{1}d_j(\theta;W)|j_F(W)\rangle,\\
&&|\widetilde\chi_2(\theta;W)\rangle=\sum_{j=0}^{1}c_j(\theta;W)|j_F(W)\rangle ,\nonumber
\end{eqnarray}
and
\begin{eqnarray}\label{normalization2}
 &&d_j(\theta;W)=\int_{-\infty}^{\infty}\varphi_j^*(V;W)\chi_W(V,\theta)\,dV,\nonumber\\
 &&\mathcal N(\theta;W)=\sqrt{|d_0|^2+|d_1|^2},\\
 &&c_j(\theta;W)=\frac{d_j(\theta;W)}{\mathcal N(\theta;W)},
 \qquad \mathcal N>0.\nonumber
\end{eqnarray}
We note that truncated state $|\widetilde\chi_2(\theta;W)\rangle$ is normalized. 
The projected fast state is then
\begin{equation}\label{2solution}
|\widetilde\chi_2(t;W)\rangle=c_0(t;W)|0_F(W)\rangle+c_1(t;W)|1_F(W)\rangle.
\end{equation}

\section{Spikes, qubit and coherence}


\begin{figure*}
\centering
\subfloat[\label{subfig1a}]{\includegraphics[width=0.5\textwidth]{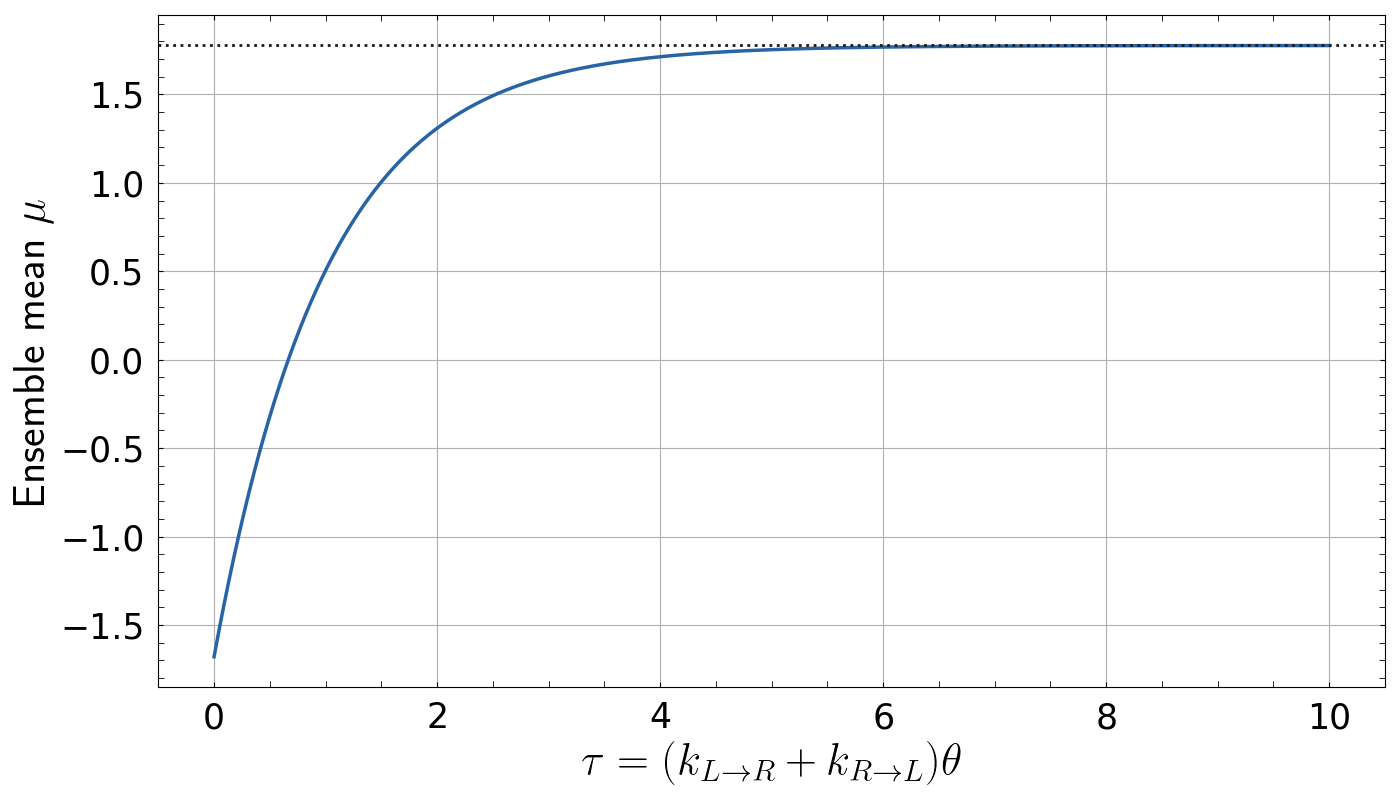}}
\hfill
\subfloat[\label{subfig2a}]{\includegraphics[width=.5\textwidth]{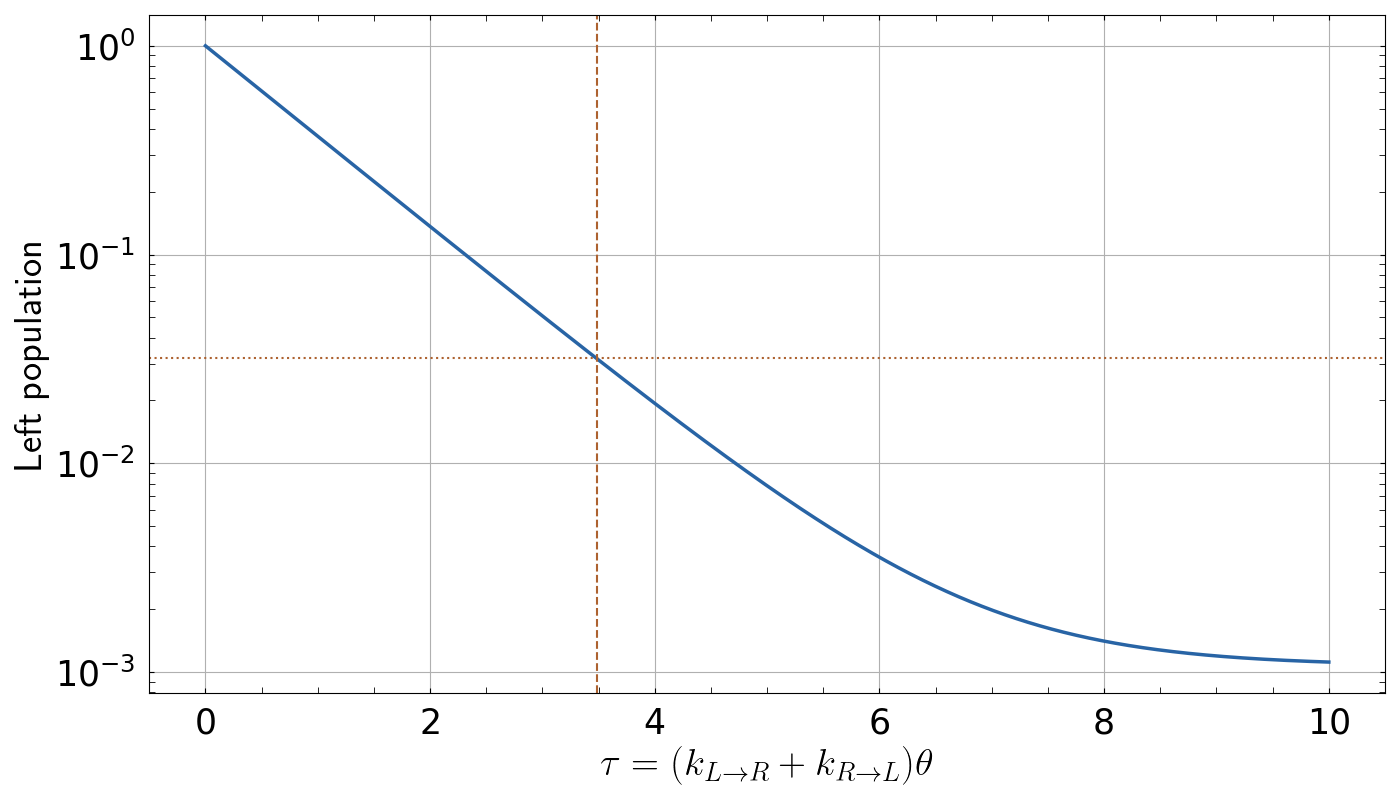}}
\caption{ Transient auxiliary coherence during fixed-$W$ relaxation in the two-state approximation. The horizontal axis is the relaxation-scaled time $\tau=k_\Sigma\theta$, where $k_\Sigma=k_{L\to R}^{(\theta)}+k_{R\to L}^{(\theta)}$. (a) The ensemble mean $\mu\simeq p_LV_L+p_RV_R$ approaches the right-well region monotonically; the horizontal dotted line marks the right minimum $V_R$. (b) The left-basin population $p_L$ on a logarithmic scale, evolving as $p_L(\tau)=p_L^{\mathrm{st}}+[p_L(0)-p_L^{\mathrm{st}}]e^{-\tau}$, with $p_L^{\mathrm{st}}=k_{R\to L}^{(\theta)}/k_\Sigma\simeq\pi_L$. The orange dotted line marks $p_L^\star=\sqrt{\pi_L}/(\sqrt{\pi_L}+\sqrt{\pi_R})$. The vertical dashed line marks the passage through $p_L=p_L^\star$, where equal normalized amplitude weights give the maximum $C_{l_2,LR}^{(\chi)}=1/\sqrt{2}$, indicated by the horizontal dotted line.}\label{ensmeble_mean}
\end{figure*}

\begin{figure}[ht]
\centering
\includegraphics[width=\linewidth]{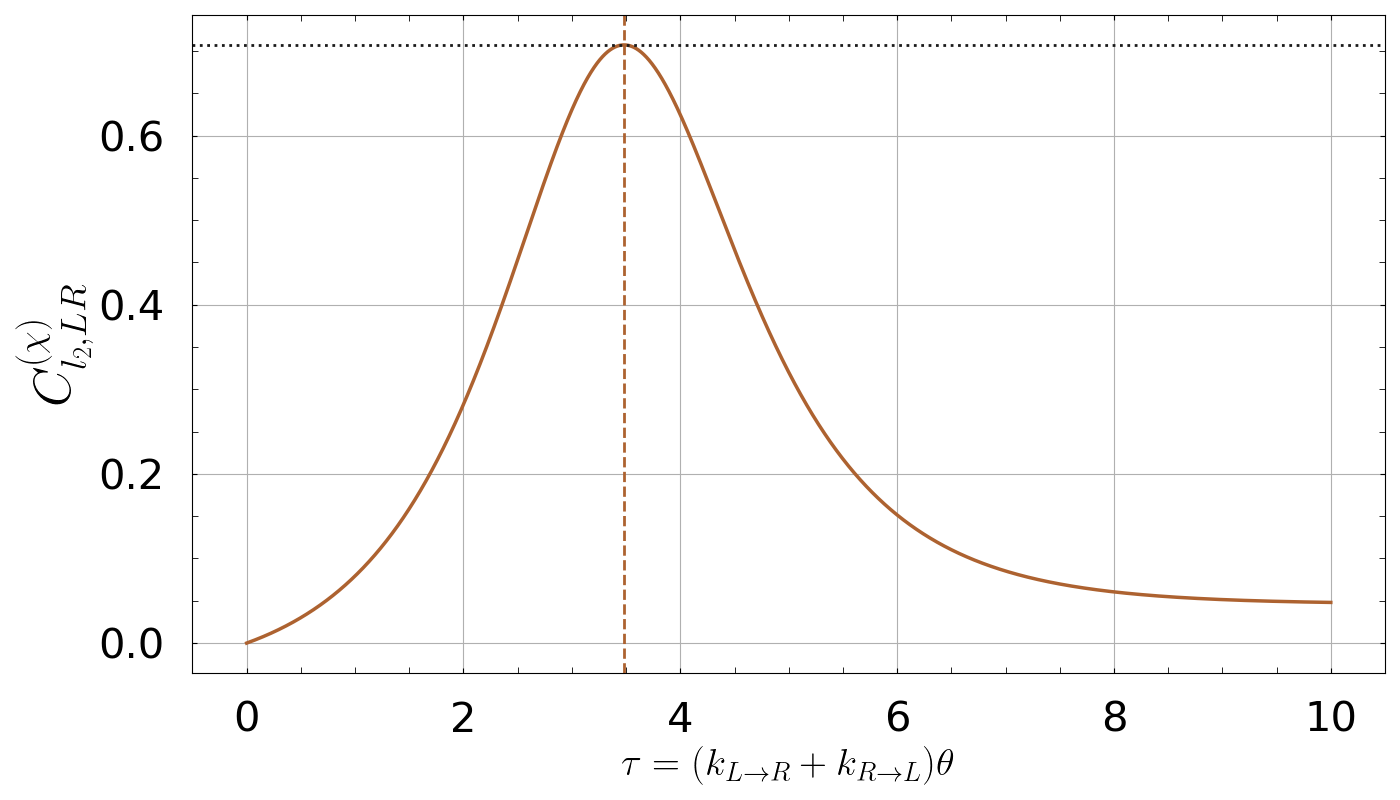}
\hfill
\caption{Transient auxiliary coherence during fixed-$W$ relaxation in the two-state approximation. The horizontal axis is the relaxation-scaled time $\tau=k_\Sigma\theta$, where $k_\Sigma=k_{L\to R}^{(\theta)}+k_{R\to L}^{(\theta)}$. the localized-basis coherence $C_{l_2,LR}^{(\chi)}\simeq
\sqrt{2\pi_L\pi_R}\,p_Lp_R/(\pi_Rp_L^2+\pi_Lp_R^2)$ of the normalized similarity-transformed amplitude. The vertical dashed line marks the passage through $p_L=p_L^\star$, where equal normalized amplitude weights give the maximum $C_{l_2,LR}^{(\chi)}=1/\sqrt{2}$, indicated by the horizontal dotted line. The subsequent relaxation approaches the small, nonzero equilibrium value $\sqrt{2\pi_L\pi_R}$. This example exhibits a coherence pulse without a spike in the ensemble mean.}\label{auxialiary_coherence}
\end{figure}

\begin{figure*}
\centering
\subfloat[\label{subfig3a}]{\includegraphics[width=0.5\textwidth]{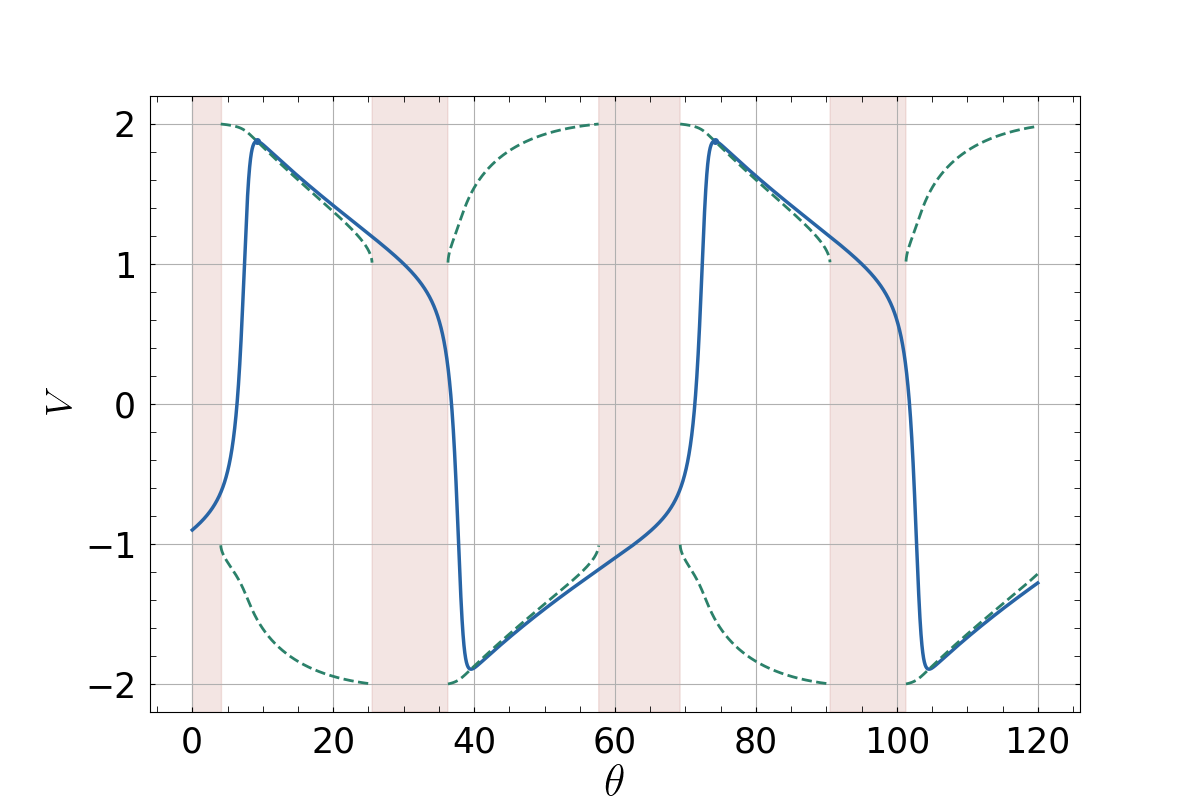}}
\hfill
\subfloat[\label{subfig3b}]{\includegraphics[width=.5\textwidth]{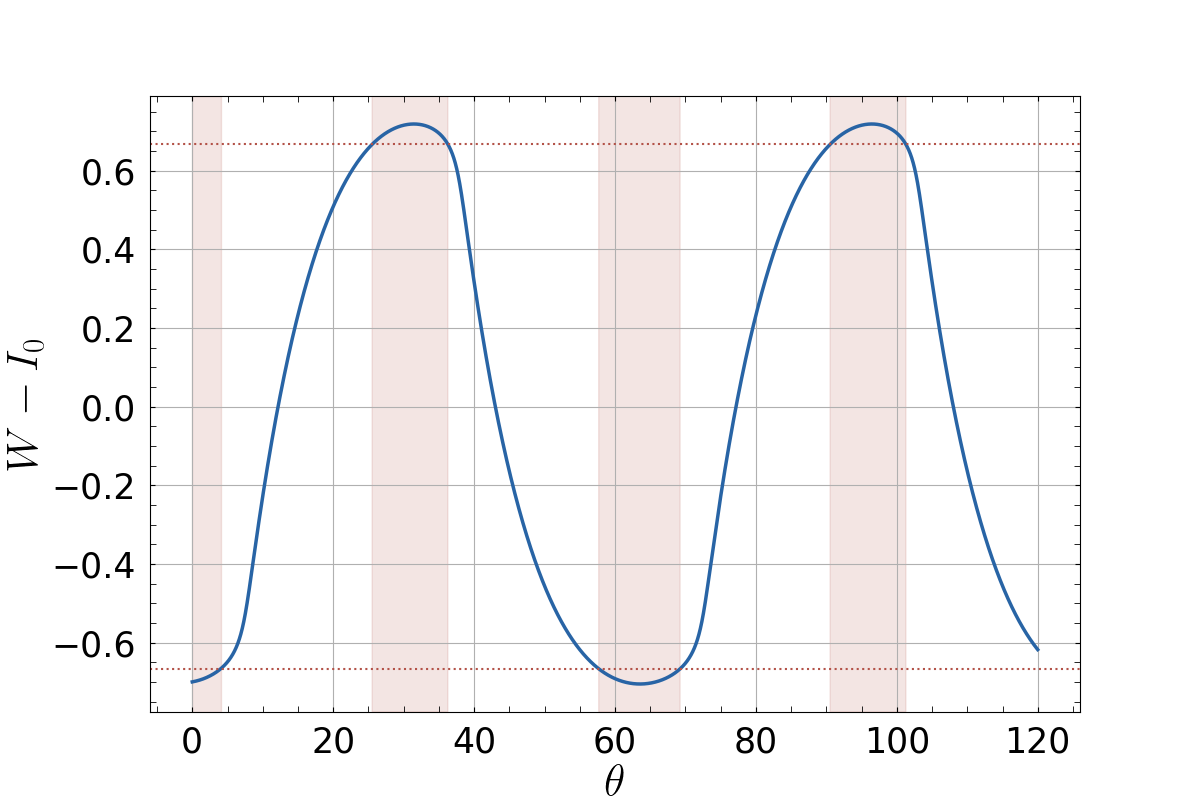}}
\caption{Validity of the frozen double-well approximation along the Case~1 deterministic FHN waveform. (a) A single deterministic activation trajectory $V(\theta)$ (solid blue), compared with the instantaneous minima $V_L$ and $V_R$ (dashed green) of $U_W(V)=V^4/12-V^2/2+(W-I_0)V$ in the bistable domain. (b) The time-dependent tilt $W(\theta)-I_0$; the horizontal dotted lines mark the folds $W-I_0=\pm2/3$. Pink shading identifies intervals outside the bistable domain, where a localized two-well description is unavailable. }\label{Frequency map}
\end{figure*}

\begin{figure}[ht]
\centering
\includegraphics[width=\linewidth]{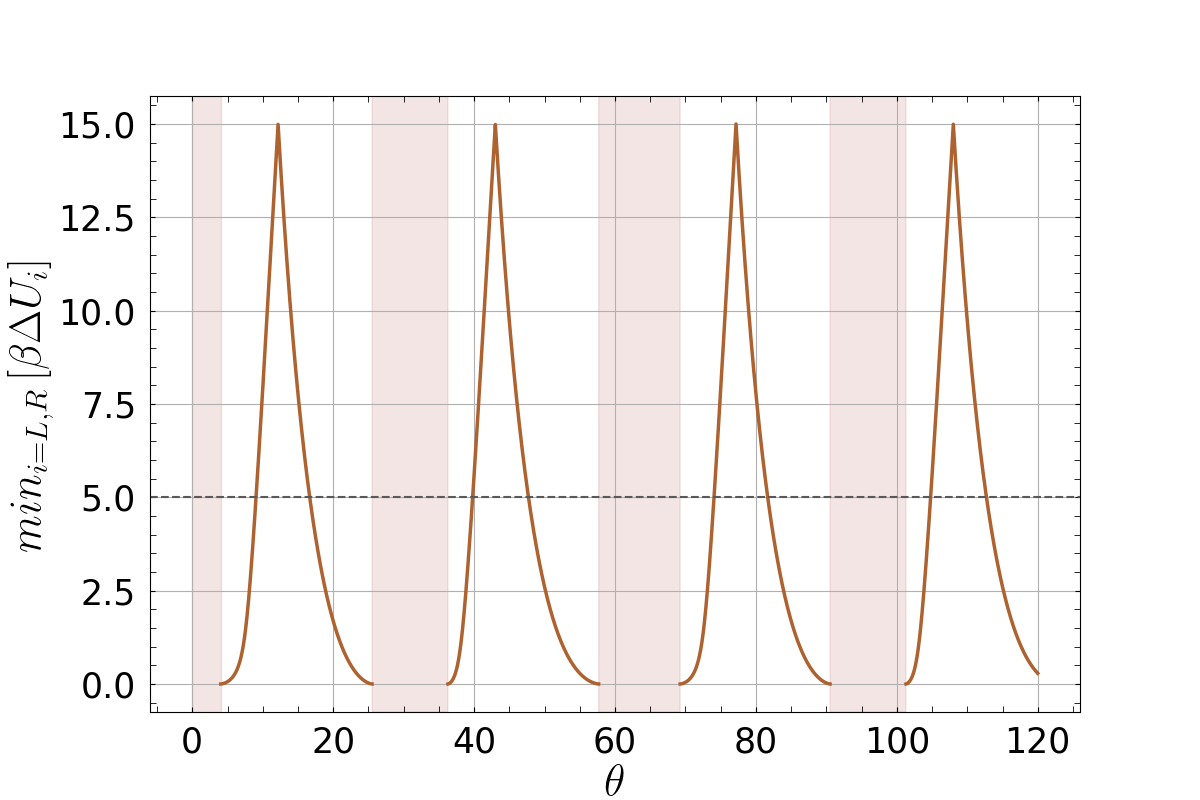}
\hfill
\caption{Validity of the frozen double-well approximation along the Case~1 deterministic FHN waveform. The smaller dimensionless escape barrier
$\min_{i=L,R}[\beta\Delta U_i]$ within the bistable intervals, with $\Delta U_i=U_W(V_S)-U_W(V_i)$ and $V_S$ the intervening barrier top. The horizontal dashed line at 5 is the screening level used in the plot; the weak-noise approximation requires both $\beta\Delta U_L\gg1$ and $\beta\Delta U_R\gg1$, together with adequate separation of time scales. Pink shading identifies intervals outside the bistable domain, where a localized two-well description is unavailable. Repeated loss of bistability and low-barrier intervals prevent uniform application of the localized two-state approximation throughout the spike cycle. The plotted $V(\theta)$ is a single trajectory, not the ensemble mean needed to infer basin populations from the mixture approximation.}\label{barrier_coherence}
\end{figure}


We construct a two-state representation from the similarity-transformed probability density
$\chi_W=e^{\beta U_W/2}P_W$. For fixed $W$ and constant input $I_0$, its evolution in the FHN time is described by
$\partial_\theta\chi_W=-D_V\widehat H_B(W)\chi_W$. This imaginary-time evolution does not generally preserve the ordinary
Hilbert-space norm. We therefore use the normalized projection $|\widetilde\chi_2\rangle$ defined above. This construction provides an auxiliary amplitude representation of classical stochastic dynamics and is distinct from the Nelson amplitude $\psi_W=\sqrt{P_W}\exp(iS_W/\widehat\hbar_{AF})$.

In the spectral basis $\{|0_F\rangle,|1_F\rangle\}$, the density matrix $\widehat\varrho^{(\chi)}
=|\widetilde\chi_2\rangle\langle\widetilde\chi_2|$ has unit trace and its Hilbert--Schmidt off-diagonal norm is
\begin{eqnarray} \label{eq:chi-eigen-coherence}
C_{l_2,\mathrm{eig}}^{(\chi)}=\sqrt{2}\,|c_0c_1|,\qquad
\bigl(C_{l_2,\mathrm{eig}}^{(\chi)}\bigr)^2=2|c_0|^2|c_1|^2.
\end{eqnarray}
This quantity depends on the chosen basis. To relate the representation to occupation of the two branches, we next specify a localized basis.

For $|W-I_0|<2/3$, let $V_L<V_S<V_R$ we denote the extrema of the same frozen potential $U_W(V)=V^4/12-V^2/2+(W-I_0)V$.
The outer roots are minima and satisfy $|V_{L/R}|>1$. We define the basins and their classical populations by
\begin{eqnarray} \label{eq:chi-basin-populations}
&&\mathcal B_L=(-\infty,V_S), \mathcal B_R=(V_S,\infty),\nonumber\\
&&p_i(\theta;W)=\int_{\mathcal B_i}P_W(V,\theta)\,dV,\\
&&p_L+p_R=1,\qquad i=L,R .\nonumber
\end{eqnarray}
The corresponding equilibrium weights are
\begin{eqnarray} \label{eq:chi-equilibrium-weights}
 &&Z_i(W)=\int_{\mathcal B_i}e^{-\beta U_W(V)}\,dV,\nonumber\\
 &&Z_W=Z_L+Z_R,\\
&&\pi_i(W)=\frac{Z_i(W)}{Z_W}.\nonumber
\end{eqnarray}
The symbols $p_i$ describe time-dependent occupations, whereas $\pi_i$ describe equilibrium occupations at the specified $W$ and noise strength. In the weak-noise regime,
$Z_i\simeq e^{-\beta U_W(V_i)}\sqrt{2\pi/(\beta\kappa_i)}$, where $\kappa_i=U_W''(V_i)>0$. An orthonormal basis within the two-mode subspace is obtained by the rotation
\begin{eqnarray}\label{eq:chi-localized-basis}
&&\varphi_L=\sqrt{\pi_L}\,\varphi_0+\sqrt{\pi_R}\,\varphi_1,\\
&&\varphi_R=\sqrt{\pi_R}\,\varphi_0-\sqrt{\pi_L}\,\varphi_1.\nonumber
\end{eqnarray}
The sign of $\varphi_1$ is chosen positive in the left well. This rotation is orthonormal exactly; its interpretation as a branch-localized basis uses the metastable, weak-noise approximation. In that regime, the local profiles reduce to the Gaussian wave packets introduced in the preceding section. The normalized projected state has the expansion
\begin{eqnarray}\label{eq:chi-normalized-local-expansion}
&&|\widetilde\chi_2\rangle=c_L|L\rangle+c_R|R\rangle,\nonumber\\
&&c_L=\sqrt{\pi_L}\,c_0+\sqrt{\pi_R}\,c_1,\\
&&c_R=\sqrt{\pi_R}\,c_0-\sqrt{\pi_L}\,c_1 ,\nonumber
\end{eqnarray}
with $|c_L|^2+|c_R|^2=1$. Its matrix in this basis is
\begin{eqnarray}\label{eq:chi-local-density-matrix}
 \varrho_{LR}^{(\chi)}
 =\begin{pmatrix}
 |c_L|^2 & c_Lc_R^*\\
 c_L^*c_R & |c_R|^2
 \end{pmatrix},
\end{eqnarray}
and
\begin{eqnarray}\label{eq:chi-local-density-matrix2}
C_{l_2,LR}^{(\chi)}=\sqrt{2}\,|c_Lc_R|.
\end{eqnarray}
The entries $|c_i|^2$ are weights of the normalized transformed amplitude. They need not equal the classical basin populations $p_i$. To derive their relation, assume rapid equilibration within each well and negligible probability in the barrier region. We write
\begin{eqnarray} \label{eq:chi-local-equilibrium-mixture}
&&P_W(V,\theta)\simeq p_L(\theta)P_L(V|W)+\nonumber\\
&&p_R(\theta)P_R(V|W),\\
&&P_i(V|W)=\frac{e^{-\beta U_W(V)}}{Z_i(W)}\,\mathbf 1_{\mathcal B_i}(V).\nonumber
\end{eqnarray}
Here $\mathbf 1_{\mathcal B_i}$ denotes the basin indicator. The functions $\sqrt{P_i}$ approximate the localized profiles away from the barrier; the smooth spectral basis is used for operator calculations. Since $\varphi_i\simeq\sqrt{P_i}$ within each well, the similarity transformation yields
\begin{eqnarray}\label{eq:chi-raw-local-amplitudes}
\chi_W\simeq\frac{p_L}{\sqrt{Z_L}}\,\varphi_L+\frac{p_R}{\sqrt{Z_R}}\,\varphi_R,\qquad
 d_i\simeq\frac{p_i}{\sqrt{Z_W\pi_i}}.
\end{eqnarray}
The expansion in Eq.~\eqref{eq:chi-raw-local-amplitudes} is unnormalized. After normalization, the common factor $Z_W^{-1/2}$ cancels and we obtain
\begin{eqnarray}\label{eq:chi-population-amplitude-relation}
&&c_i\simeq\frac{p_i/\sqrt{\pi_i}}{\sqrt{p_L^2/\pi_L+p_R^2/\pi_R}},\\
&&|c_i|^2\simeq\frac{p_i^2/\pi_i}{p_L^2/\pi_L+p_R^2/\pi_R}.\nonumber
\end{eqnarray}
Consequently, the localized-basis coherence of the normalized similarity-transformed state reads
\begin{widetext}
\begin{eqnarray}\label{eq:chi-general-coherence}
C_{l_2,LR}^{(\chi)}(\theta;W)\simeq\frac{\sqrt{2\pi_L(W)\pi_R(W)}\,p_L(\theta;W)p_R(\theta;W)}{\pi_R(W)p_L^2(\theta;W)+\pi_L(W)p_R^2(\theta;W)} .
\end{eqnarray}
\end{widetext}
This expression replaces the identification $C_{l_2}=\sqrt{2p_Lp_R}$ for the $\chi_W$ representation. The latter formula would describe a separately defined square-root population encoding, rather than the projection of $e^{\beta U_W/2}P_W$. For the symmetric potential, $W=I_0$ and $\pi_L=\pi_R=1/2$, so that
\begin{eqnarray}\label{eq:chi-symmetric-coherence}
C_{l_2,LR}^{(\chi)}\simeq\frac{\sqrt{2}\,p_Lp_R}{p_L^2+p_R^2}. 
\end{eqnarray}
In general, $0\leq C_{l_2,LR}^{(\chi)}\leq1/\sqrt{2}$. The maximum corresponds to equal normalized amplitude weights and, within Eq.~\eqref{eq:chi-population-amplitude-relation}, occurs at
\begin{eqnarray}\label{eq:chi-maximum-condition}
p_L=\frac{\sqrt{\pi_L}}{\sqrt{\pi_L}+\sqrt{\pi_R}},\qquad
p_R=\frac{\sqrt{\pi_R}}{\sqrt{\pi_L}+\sqrt{\pi_R}}.
\end{eqnarray}
Equal classical populations maximize the coherence only in the
symmetric case. At equilibrium, $p_i=\pi_i$, and
\begin{eqnarray}\label{eq:chi-equilibrium-coherence}
C_{l_2,LR,\mathrm{eq}}^{(\chi)}=\sqrt{2\pi_L\pi_R},\qquad
C_{l_2,\mathrm{eig},\mathrm{eq}}^{(\chi)}=0.
\end{eqnarray}
The second equality follows because the normalized equilibrium state is $\varphi_0$. These different values demonstrate the basis dependence of the diagnostic. In particular, small noise alone does not imply small localized-basis coherence: the symmetric equilibrium state has $C_{l_2,LR}^{(\chi)}=1/\sqrt{2}$. The time dependence follows from the probability dynamics. For a fixed $W$, the two-state approximation gives
\begin{eqnarray} \label{eq:chi-occupation-kinetics}
\frac{dp_L}{d\theta}=-k_{L\to R}^{(\theta)}p_L+k_{R\to L}^{(\theta)}p_R,\qquad p_R=1-p_L .
\end{eqnarray}
With $k_\Sigma=k_{L\to R}^{(\theta)}+k_{R\to L}^{(\theta)}$,
\begin{eqnarray} \label{eq:chi-frozen-population-solution}
&&p_L(\theta)=p_L^{\mathrm{st}}+[p_L(0)-p_L^{\mathrm{st}}]e^{-k_\Sigma\theta},\nonumber\\
&&p_L^{\mathrm{st}}=\frac{k_{R\to L}^{(\theta)}}{k_\Sigma}\simeq\pi_L .
\end{eqnarray}
The rate $k_\Sigma$ approximates the first nonzero Fokker--Planck relaxation eigenvalue $\lambda_1^{(\theta)}$ obtained from the supersymmetric construction. Equations
Eq.\eqref{eq:chi-general-coherence} and Eq.\eqref{eq:chi-occupation-kinetics} therefore connect the spectral relaxation rate to the evolution of the auxiliary two-state coherence.

An alternative expression can be obtained from the ensemble mean. With $\mu(\theta;W)=\langle V\rangle_W$ and narrow intrawell distributions:
\begin{eqnarray} \label{eq:chi-mean-to-populations}
&&\mu\simeq p_LV_L+p_RV_R,\nonumber\\
&&p_L\simeq\frac{V_R-\mu}{V_R-V_L},\\
&&p_R\simeq\frac{\mu-V_L}{V_R-V_L}.\nonumber
\end{eqnarray}
Substitution gives
\begin{eqnarray} \label{eq:chi-mean-coherence}
C_{l_2,LR}^{(\chi)}\simeq\frac{\sqrt{2\pi_L\pi_R}\,(V_R-\mu)(\mu-V_L)}{\pi_R(V_R-\mu)^2+\pi_L(\mu-V_L)^2}.
\end{eqnarray}
This formula requires an ensemble mean satisfying $V_L\leq\mu\leq V_R$ within the mixture approximation. It does not determine branch probabilities from a single FHN
trajectory. Beyond the narrow-well limit, the conditional means $\int V P_i(V|W)\,dV$ replace the minima $V_i$ in Eq.~\eqref{eq:chi-mean-to-populations}.

When the recovery variable varies slowly, the populations may be propagated with instantaneous rates and equilibrium weights, provided intrawell equilibration is fast relative to the change of the wells. This is an additional adiabatic approximation; the fixed-$W$ exponential solution is not used unchanged for $W=W(\theta)$. Recovery noise must also be small enough for this frozen-parameter description to apply. The localized two-state formulas require well-separated modes and $\beta\Delta U_{L/R}\gg1$; they are not uniformly valid near the folds $|W-I_0|=2/3$ or throughout an arbitrary fast spike excursion. Thus coherence peaks during spiking must be established from the evolving distribution and a validity check, rather than assumed from the deterministic waveform. The quantity $C_{l_2,LR}^{(\chi)}$ characterizes the chosen normalized amplitude representation and does not, by itself, establish physical quantum coherence in the magnetic device.


\section{Conclusions}

We have developed a supersymmetric spectral framework for the stochastic
FitzHugh--Nagumo model obtained through a local reduction of the adopted
antiferromagnetic tangent-torque dynamics. The construction links the
magnetic control parameters to an effective slow--fast stochastic system.
Because the full two-dimensional FHN drift is non-gradient, the scalar
factorization used here is applied to the frozen fast-coordinate problem.
For fixed recovery coordinate and constant input, the resulting
one-dimensional Fokker--Planck generator is related by a similarity
transformation to a positive semidefinite supersymmetric Hamiltonian.

The first nonzero relaxation eigenvalue of this Hamiltonian equals the
ground-state eigenvalue of its supersymmetric partner. In the weak-noise
bistable regime, the corresponding relaxation rate in the original FHN
time $\theta$ is
\begin{eqnarray}
&&\lambda_1^{(\theta)}(W)\simeq k_{L\to R}^{(\theta)}(W)+k_{R\to L}^{(\theta)}(W)
\simeq\frac{\sqrt{\kappa_L\kappa_S}}{2\pi}e^{-\beta\Delta U_L}\nonumber\\
&&+\frac{\sqrt{\kappa_R\kappa_S}}{2\pi}e^{-\beta\Delta U_R}.
\end{eqnarray}

Here $\kappa_{L,R}$ and $\kappa_S$ are the well and barrier curvatures,
and $\Delta U_{L,R}$ are the escape barriers. The eigenvalue in the
rescaled time $t=D_V\theta$ is
$\lambda_1^{(t)}=\beta\lambda_1^{(\theta)}$, with
$D_V=1/\beta$. This relation provides a spectral characterization of
interwell equilibration and of the rates entering the two-state population
dynamics.

A normalized projection of $\chi_W=e^{\beta U_W/2}P_W$ onto the two
lowest modes defines an auxiliary two-state amplitude. Its coefficients
in the localized basis are not the square roots of the classical basin
populations. Under the metastable mixture approximation,
$c_i\propto p_i/\sqrt{\pi_i}$, and the Hilbert--Schmidt off-diagonal norm is
\begin{equation*}
C_{l_2,LR}^{(\chi)}(\theta;W)
\simeq
\frac{\sqrt{2\pi_L\pi_R}\,p_Lp_R}
     {\pi_Rp_L^2+\pi_Lp_R^2}.
\end{equation*}
The upper bound $1/\sqrt{2}$ is reached when
$p_L/p_R=\sqrt{\pi_L/\pi_R}$; equal basin populations attain this bound
only for symmetric equilibrium weights. At equilibrium, the localized-basis
value is $\sqrt{2\pi_L\pi_R}$, whereas the spectral-basis value vanishes.
Thus both the amplitude representation and the basis must be specified
when interpreting this diagnostic.

The illustrative examples clarify its relation to spiking. In the fixed-$W$ example,
monotonic population relaxation produces a transient coherence maximum
while the ensemble mean approaches its stationary value without a spike.
A coherence pulse therefore need not identify an excitable excursion.
Conversely, the illustrated Case~1 spiking waveform repeatedly crosses the
folds $|W-I_0|=2/3$ and includes bistable intervals with small escape
barriers. The localized two-state description consequently does not apply
uniformly along the full cycle. A deterministic waveform alone cannot
supply the ensemble populations required to evaluate the coherence.

The framework therefore connects relaxation spectra, population transfer,
and an auxiliary coherence diagnostic within a stated regime of validity.
Its extension to an evolving recovery coordinate requires sufficiently
rapid intrawell equilibration, weak recovery noise, and well-separated
metastable modes. Direct propagation of the stochastic distribution is
needed to assess coherence during complete spike cycles. The resulting
two-state construction represents classical stochastic dynamics; evidence
for physical quantum coherence in an antiferromagnetic device would require
additional microscopic analysis and observables.

\begin{acknowledgments}
The author D. Maroulakos gratefully acknowledges support from the Doctoral School at
the University of Rzesz\'ow and the NAWA STER scholarship program for the best foreign doctoral students studying at the Doctoral School at the University of Rzesz\'ow within the framework of the Project ``UR an international PhD student'' (Decision of the Director of NAWA No. BPI/STE/2023/1/00001/DEC/01 dated 19.10.2023) in accordance with the assumptions of the NAWA STER Program - Internationalization of Doctoral Schools.
\end{acknowledgments}

\bibliography{FitzHugh.Nagumo}

\end{document}